\documentstyle[prl,aps,preprint,floats]{revtex}

\newcommand{\mn}{{\mu\nu}}

\newcommand{\be}{\begin{equation}}
\newcommand{\ee}{\end{equation}}
\newcommand{\half}{\frac{1}{2}}

\def\PR{Phys. Rev. }
\def\ZP{Z. Phys. C}
\def\PL{Phys. Lett.}  
\def\PRL{Phys. Rev. Lett.}  
\def\NP{{ Nucl. Phys. }}

\def\J{$J/\psi$}
\def\j{J/\psi}

\def\e{\epsilon}
\def\c2{$\chi_2$}

\begin{document}
\draft

\setcounter{page}{0}


\preprint{\vbox{MIT-CTP-2838 and RIKEN-BNL preprint \null\hfill\rm March 5, 1999}}


\title{$\mathbf{\chi_2}$ production in polarized $\mathbf{pp}$ collisions at
   RHIC: \\ measuring $\mathbf{\Delta G}$ and testing the color octet model}

\author{R.L.~Jaffe$^a$ and D.~Kharzeev$^b$\\ \null}

\address{a) Center for Theoretical Physics,\\
   Laboratory for Nuclear Science
   and Department of Physics,\\
   Massachusetts Institute of Technology,\\
   Cambridge, Massachusetts 02139\\[6mm]
   b) RIKEN-BNL Research Center,\\
   Brookhaven National Laboratory,\\
Upton, New York 11973\\}

\maketitle

\begin{abstract}

\noindent We consider the production and decay of the $\chi_2$ 
charmonium state in polarized and unpolarized $pp$ collisions at RHIC 
in the framework of an effective theory based on the QCD multipole 
expansion.  We find that the angular distribution in the decay of the 
produced charmonium, $\chi_2 \to J/\psi + \gamma$, in the unpolarized 
case allows us to distinguish clearly between the color singlet and 
color octet production mechanisms.  Once the production mechanism is 
known, the angular distribution in the polarized case can be used to 
measure the polarized gluon distribution in the proton , $\Delta 
G(x)$.
\end{abstract}


\narrowtext \newpage The production of heavy quarkonia recently has 
attracted much interest.  Tevatron data \cite{TeV} have clearly shown 
that the traditional color singlet model \cite{csm} fails to explain 
the measured $J/\psi$ and $\psi'$ production cross sections, and have 
stimulated the development of a more general approach based on the 
non-relativistic expansion in QCD \cite{bbl}.  This approach 
systematically classifies, in powers of heavy quark velocity, the 
contributions of intermediate color singlet and color octet 
quark--antiquark states to quarkonium production.  The corresponding 
long--distance color octet matrix elements unfortunately cannot be 
computed analytically at present, and have to be either extracted from 
the data or computed on the lattice.

In this Letter, we use the ideas of QCD multipole expansion \cite{mult} to
formulate a simple effective approach to treat both color singlet and
color octet intermediate states in quarkonium production.
Specifically, we consider the production of $\chi_2$ charmonium state
in polarized $pp$ collisions at small $P_t$.  This process has been
considered previously by several authors \cite{ioffe,cp,asym}.  Here we show
that the measurement of the angular distribution in the decay of the
produced charmonium, $\chi_2 \to J/\psi + \gamma$, can be used {\it
both\/} to distinguish between color octet and singlet production
mechanisms {\it and\/} to measure the polarized gluon distribution in
the polarized nucleon.  The knowledge of the production mechanism is a
necessary pre-requisite for the extraction of $\Delta G(x,Q^{2})$ from
the spin asymmetries in charmonium production.  It will also eliminate
many uncertainties that presently complicate interpretations of the
data on charmonium production in nuclear collisions \cite{rev}.

We begin with a summary of our results.  The underlying mechanism at 
the parton level is gluon-gluon fusion.  First we consider $\chi_{2}$ 
production in {\it unpolarized\/} $pp$ collisions.  The $\chi_{2}$ may 
be produced directly -- the color singlet mechanism -- with amplitude 
${\cal A}^{\mathbf{1}}$.  As pointed out in Ref.~\cite{brodskyetal}, 
direct production at low $p_{\perp}$ produces $\chi_{2}$'s with 
helicity $\pm 2$ only.  Alternatively, the $\chi_{2}$ can be produced 
in a two step process involving a color octet intermediate state.  
First two gluons annihilate to form a $c\bar c$ pair in a color 
$\mathbf{8}$ state.  As we argue below, the $J^{\rm PC} = 1^{--}$ 
$c\bar c$ configuration dominates.  This decays by a color E1 
transition to the $\chi_{2}$ with helicity $\pm 1$ only.  We 
parameterize the production and decay process by an amplitude ${\cal 
A}^{\mathbf{8}}$.  ${\cal A}^{\mathbf{1}}$ and ${\cal A}^{\mathbf{8}}$ 
are independent of the kinematic variables $x_{1}$, $x_{2}$ and 
$\theta$.  The angular distribution of the photon relative to the beam 
axis in the decay $\chi_{2}\to J/\psi +\gamma$ determines the ratio of 
${\cal A}^{\mathbf{8}}$ to ${\cal A}^{\mathbf{1}}$,
\begin{equation}
    \frac{d\sigma}{d\Omega} \propto
    g(x_{1},M_{\chi}^{2})g(x_{2},M_{\chi}^{2}) \left\{\left(\frac{1}{2} + \half
    \cos^{2}\theta\right){\cal A}^{\mathbf{1}}+ \left(\frac{3}{4} - \frac{1}{4}
    \cos^{2}\theta\right){\cal A}^{\mathbf{8}}\right\}
    \label{unpol}
\end{equation}

Once ${\cal A}^{\mathbf{8}}/{\cal A}^{\mathbf{1}}$ has been measured 
in unpolarized $pp$ collisions, the same process can be used to 
measure $\Delta g(x, M_{\chi}^{2})$ at a polarized $pp$ collider like 
RHIC. The angular distribution of the spin asymmetry in $\chi_{2}$ 
production is given by,
\begin{equation}
    \frac{d\sigma^{\uparrow\uparrow} -
    d\sigma^{\uparrow\downarrow}} {d\sigma^{\uparrow\uparrow} +
    d\sigma^{\uparrow\downarrow}} = -\frac{\Delta g (x_{1},
    M_{\chi}^{2})}{ g (x_{1}, M_{\chi}^{2})}
    \frac{\Delta g (x_{2}, M_{\chi}^{2})}{ g (x_{2}, M_{\chi}^{2})}
    \times \frac {\frac{1}{2}+\half\cos^{2}\theta -
     \frac{{\cal A}^{\mathbf{8}}}{{\cal A}^{\mathbf{1}}}
     \{\frac{3}{4} - \frac{1}{4}\cos^{2}\theta\}}
     {\frac{1}{2}+\half\cos^{2}\theta +
     \frac{{\cal A}^{\mathbf{8}}}{{\cal A}^{\mathbf{1}}}
     \{\frac{3}{4} - \frac{1}{4}\cos^{2}\theta\}}
    \label{ang}
\end{equation}

Equations (\ref{unpol}) and (\ref{ang}) are our basic results.  
After a brief discussion of the restrictions on their 
validity, we present a derivation. We assume that the 
$\chi_{2}$ is produced by gluon-gluon fusion.  Light quark fusion is 
known to become important only at $x_{F}>0.4$ \cite{gav}.  We ignore 
the gluon transverse momentum.  It is easy to show that intrinsic 
transverse momentum generates corrections of order 
$k_{T}^{2}/M_{\chi}^{2}$.  The most important $k_{T}$ 
correction is the possibility to produce the $\chi_2$ with 
$\Lambda=0$, with amplitude $\sim k_T^2/M^2$.  $k_T$ effects can be 
eliminated altogether experimentally at some cost in rate, by limiting 
the total transverse momentum of the photon and dilepton to a small 
value. 

We work to lowest non-trivial order in perturbative 
QCD. Hard QCD corrections to charmonium production by gluon-gluon 
fusion have been studied in Ref.~\cite{nlo}.  They can be ignored for 
$\chi_{2}$'s produced at low $k_{T}$.  Our results follow from three 
observations:
\begin{enumerate}
    \item Direct production of the $\chi_{2}$ in the color singlet
    state occurs only in the helicity $\pm 2$ states.
    \item Production of the $\chi_{2}$ through an intermediate color
    octet state yields $\chi_{2}$'s predominantly with helicity $\pm 1$.
    \item Each $\chi_{2}$ helicity state decays into $\j +\gamma$ with
    a characteristic angular distribution.
\end{enumerate}
Although (2) is the only ingredient not either in the literature or 
obtainable by elementary means, for completeness we review all three.

We describe the \c2 state of charmonium by a composite tensor field
$\eta^{\mn}$, which is symmetric, $\eta^{\mn}=\eta^{\nu\mu}$,
traceless, $\eta^{\mu}_{\mu}=0$, and transverse,
$\partial_{\mu}\eta^{\mn}=0$.  It is convenient to define the \c2\
wave function $H^{\mn}(P, \Lambda)$ with fixed momentum $P$ and
helicity $\Lambda$:
\be
        \langle P, \Lambda | \eta^{\mn} | 0 \rangle = H^{\mn}(P,
        \Lambda)\quad\hbox {with}\quad H^{\mn}=H^{\nu\mu},\quad
        H^{\mu}_{\mu}=0,\quad P_{\mu} H^{\mn} =0.
        \label{wf}
\ee
In the rest frame of the \c2, the transversality condition implies that $
H^{00} = H^{0j} = H^{j0} = 0$.  The wave function in the rest frame
therefore reduces as expected to an irreducible 3--tensor of rank 2,
with $5$ independent components describing different spin projections.

We must consider two couplings of the \c2\ state to gauge vector 
bosons.  First, the electromagnetic decay coupling to the photon plus 
\J\ and second, the QCD production coupling to a $[1^{--}]^{\mathbf 
8}$ ($c\bar c$) state and a gluon.  Both are electric dipole 
transitions with effective interaction Lagrangians,
 \begin{eqnarray}
     {\cal{L}}_{\rm I} &=& g_{\rm I}\,\psi_{\mu}^{\dagger} D_{\nu}
     \eta^{\mn} \ \sim \ \psi_{\mu}^{\dagger} A_{\nu}
     \eta^{\mn},\nonumber\\
     {\cal{L}}_{\rm II} &=& g_{\rm II}\,\hbox
     {Tr}\,\{\mathbf{\Psi}_{\mu}^{\dagger} \mathbf{D}_{\nu}
     \}\,\eta^{\mn} \ \sim \ \hbox{Tr}\,\{\mathbf{\Psi}_{\mu}^{\dagger}
     \mathbf{A}_{\nu}\} \,\eta^{\mn}.
     \label{int}
 \end{eqnarray}
where $\psi_{\mu}$ ($\mathbf{\Psi}_{\mu}$) is an interpolating field 
for the \J\ ($[1^{--}]^{\mathbf 8}$ ($c\bar c$)) and $A_{\mu}$ 
($\mathbf{A}_{\mu}$) is the photon (gluon) vector potential.  
$\mathbf{A}_{\mu}$ and $\mathbf{\Psi}_{\mu}$ are matrices in the octet 
representation of SU(3)$_{\rm color}$, and Tr denotes a trace over 
color indices.  The effective Lagrangians eq.~(\ref{int}) correspond 
to the lowest order in the multipole expansion and are valid when the 
momentum $p$ of the gauge boson (photon or gluon) is small compared to 
the inverse size of charmonium $R^{-1}$: $p<<R^{-1}$.

From eq.~(\ref{int}) it is straightforward to evaluate the angular 
distributions in the decay $\chi_2 \to \j + \gamma$ when \c2 is 
produced in different helicity states.  It is convenient to perform 
the calculation in the \c2 rest frame where the amplitude 
corresponding to eq.~(\ref{int}) can be written as,
\be
{\cal A} \propto \e^{k \dagger}(\lambda_{\psi})
\e^{l \dagger}(\lambda_{\gamma})
H^{kl}(\Lambda),
\ee
where $\e$'s are the polarization vectors of $\j$ and $\gamma$.
Using the completeness relations,
\be
\sum_{\lambda_{\psi}} \e^m(\lambda_{\psi}) \e^{k \dagger}(\lambda_{\psi}) =
\delta^{km}, \quad
\sum_{\lambda_{\gamma}} \e^n(\lambda_{\gamma}) \e^{l
\dagger}(\lambda_{\gamma}) =
\delta^{nl} - \hat{k}^n\hat{k}^l;
\label{sum}
\ee
for the polarization vectors, where $\hat{k}$ is the unit vector in
the direction of the photon's momentum, the cross section can be
written down as
\be
\frac{d\sigma}{d\Omega} \propto
\sum_{\lambda_{\psi},\lambda_{\gamma}} {\cal A}^{\dagger} {\cal A}
\propto H^{\dagger
kl}(\Lambda) (\delta^{ln} - \hat{k}^l \hat{k}^n) H^{kn}(\Lambda).
\label{sigma}
\ee
The evaluation of eq.~(\ref{sigma}) simplifies if one parameterizes
$H^{jk}$ in terms of the unit vectors $\vec{\rm v}(\mu)$ that 
describe the angular wavefunction of a state with $J=1$ and 
$M_{J}=\mu$,
\be
H^{jk}(\Lambda) = \sum_{\mu,\mu'} (1 \mu 1 \mu' | 2 \Lambda)
{\rm v}^j(\mu) {\rm v}^k(\mu'). \label{clebsch}
\ee
The vector $\vec{\rm v}(\mu)$, can be represented in
 a Cartesian coordinate system as
\begin{equation}
\vec{\rm v}(\pm 1) = \mp \frac{1}{\sqrt{2}}\ (\hat e_1 \pm i\hat e_2),
\quad \vec{\rm v}(0) = \hat e_3. \label{deck}
\end{equation}
The phase convention in eq.~(\ref{deck}) corresponds to the familiar 
Cartesian representation of vector spherical 
harmonics; the $\vec{\rm v}(\mu)$ are normalized to unity, $|\vec {\rm 
v}(\mu)|^{2}=1$.  The decomposition of a vector $\vec w$ with respect 
to this basis is
\begin{equation}
    \vec w = \cos\theta \ \vec{\rm v}(0) +\frac{1}{\sqrt{2}}\sin\theta\ 
e^{i\phi}\ \vec{\rm 
    v}(-1) - \frac{1}{\sqrt{2}}\sin\theta\ e^{-i\phi}\ \vec{\rm v}(1) 
\label{spher}
\end{equation}
It is convenient to choose $\theta$ as the polar angle 
relative to the spin quantization axis (which we 
identify with the hadron beam axis); the polar angle $\phi$ does not 
play any role in our case.

The relations eq.~(\ref{sum},\ref{clebsch},\ref{deck},\ref{spher})
now make the evaluation of the angular dependence of the cross section
(\ref{sigma}) trivial. For different projections of the \c2's spin,
we find
\begin{eqnarray}
W^{2,\pm 2} (\theta) &\propto& \frac{1}{2} + \frac{1}{2}
\cos^2\theta,\nonumber\\
W^{2,\pm 1} (\theta) &\propto& \frac{3}{4} - \frac{1}{4}
\cos^2\theta,\nonumber\\
W^{2,0} (\theta) &\propto&  \frac{5}{6} - \frac{1}{2} \cos^2\theta.
\label{dist}
\end{eqnarray}
These decay distributions are rather distinct, a fact which can be
used for experimental determination of the helicity state of the \c2.

Let us now turn to the mechanism of \c2 production.  We are concerned 
with two competing mechanisms -- ``color singlet'' and ``color octet'' 
production.  In the first, two gluons annihilate directly to the \c2\ 
-- a color ${\mathbf 1}$ $c\bar c$ state with $J^{\rm PC}=2^{++}$.  In 
the second, two gluons annihilate first to a color ${\mathbf 8}$ 
$c\bar c$ state which then decays to the \c2\ .

The helicity dependence of the color singlet mechanism was studied in 
\cite{brodskyetal}, who concluded that the \c2\ is produced only in 
the helicity $\pm 2$ state.  Helicity $0$ is allowed by conservation 
laws but does not occur.  We agree with this result.  Our calculation 
proceeds as follows.  First we compute the amplitude, ${\cal M}$, 
coupling a \c2\ at rest with helicity $\Lambda$ to a non-relativistic 
quark antiquark pair parameterized by a relative momentum $\vec q$ and 
two component spinors, $U$ and $V$.  The momentum space wavefunction 
of the \c2\ , $\phi(q)$, need not be specified except that it falls 
rapidly with $q=|\vec q|$,
\begin{equation}
    {\cal M}(\Lambda, \vec q, U, V) =
    \phi(q)\sum_{\mu\mu'}(1\mu1\mu'|2\Lambda) (\vec {\rm v}(\mu)\!\cdot\!\vec
    q )(\vec {\rm v}(\mu')\!\cdot \!V^{\dagger}\vec\sigma U)
    \label{wavefunction}
\end{equation}
The $\vec\sigma$ and $\vec q$ in eq.~(\ref{wavefunction}) reflect the 
fact that the \c2\ state is a spin triplet $p$-wave excitation of the 
$c\bar c$ pair. 

The $c\bar c$ pair produced by gluon-gluon fusion have a distribution 
in spin and momentum determined by the elementary tree diagrams for 
$gg\to c\bar c$.  Because this process involves gluon momenta $p$ of 
the order of the charm quark mass, $p \sim M_c$, we are justified in 
describing the annihilation process to lowest non-trivial order in 
QCD. From the Feynman diagrams it is straightforward to compute the 
$\vec q$ and $\vec \sigma$ dependence of the $gg\to c\bar c$ 
amplitude, ${\cal N}$, (to lowest non-trivial order in $\vec q$) as a 
function of the gluon polarizations,
\begin{equation}
    {\cal N}(\vec q, \vec\epsilon_{1}, \vec\epsilon_{2}, U, V) \propto
    U^{\dagger}\left[ \vec\epsilon_{1}\cdot\vec q\  
    \vec\epsilon_{2}\cdot \vec\sigma + \vec\epsilon_{2}\cdot\vec q \ 
    \vec\epsilon_{1}\cdot \vec\sigma -
    \vec\epsilon_{1}\cdot\vec\epsilon_{2}\ 
    \vec q\cdot\vec\sigma + \vec\epsilon_{1}\cdot\vec\epsilon_{2}\ 
    \vec q \cdot \hat k_{1} \ \vec\sigma\cdot\hat k_{1}\right] V,
    \label{gg}
\end{equation}
where we have dropped terms zeroth order in $\vec q$ which vanish 
when averaged over the wavefunction of eq.~(\ref{wavefunction}). 
$\vec\epsilon_{1}$ and $\vec\epsilon_{2}$ are the polarization vectors for 
the incoming gluons with momenta $\vec k_{1}=-\vec k_{2}$ in the 
center of mass; $\hat{k_1}$ is a unit vector in the 
direction of $\vec k_1$.

The helicity amplitudes for $gg\to \chi_{2}$ are determined by 
combining eqs.~(\ref{wavefunction}) and (\ref{gg}), summing over quark 
spins, and integrating over $\vec q$.  Since we are interested only in 
the relative amplitudes for different helicities we ignore the 
integral over the magnitude of $\vec q$,
\begin{equation}
    {\cal R}(\Lambda,\vec\epsilon_{1},\vec\epsilon_{2}) \propto \int 
    \frac{d\Omega_{q}}{4\pi} \sum_{U,V} {\cal M}^{\dagger}(\Lambda, \vec q, U, V)
    {\cal N}(\vec q, \vec\epsilon_{1}, \vec\epsilon_{2}, U, V)
    \label{amplitude}
\end{equation}
From this it is straightforward to see that ${\cal R}$ vanishes when 
$\Lambda = 0$.  Simple helicity conservation and symmetry arguments 
then suffice to show that the amplitudes for gluons of {\it opposite} 
helicity to form a \c2\ with helicity $\pm 2$ are equal (up to an 
inconsequential phase).  This completes our confirmation of point (1) 
in our list. 

The discussion of color octet production proves simpler than color 
singlet production, because the helicity amplitudes are all determined 
by symmetry considerations once the dynamical mechanism is clear.  The 
first step is the annihilation of two gluons into a color ${\mathbf 
8}$ $c\bar c$-state.  The $(c\bar c)^{\mathbf 8}$ state can have a 
variety of J$^{\rm PC}$ quantum numbers.  It then decays to the \c2\ 
by gluon emission.  If the relative momentum of the $(c\bar 
c)^{\mathbf 8}$ pair $q$ is large compared to the inverse radius of 
quarkonium $R^{-1}$, the formation of a bound state is very unlikely.  
Formation of quarkonium therefore requires a small invariant mass of 
the pair, and in this case the momentum of the radiated gluon $p$ is 
typically small compared to $R^{-1}$, so the QCD multipole expansion 
in powers of $p R$ should be meaningful and convergent.
These considerations strongly
favor electric dipole gluon emission for the decay $(c\bar c)^{\mathbf 
8}\to \chi_{2} g$.  We are forced to conclude that the parent $(c\bar 
c)^{\mathbf 8}$ state has J$^{\rm PC}= 1^{--}$.  The only other state 
that could couple by E1 radiation would be $3^{--}$.  The $3^{--}$ is 
a $d$-wave internal excitation of the $c\bar c$ pair as opposed to the 
$1^{--}$ state, which is an s-wave.

Having established that the parent color-${\mathbf 8}$ state is a 
$1^{--}$, we can use ${\cal L}_{\rm I}$ and ${\cal L}_{\rm II}$ of 
eq.~(\ref{int}) to extract the relevant helicity amplitudes.  The 
logic is simple: The $[1^{--}]^{\mathbf 8}$ has helicity states $\pm 
1$ and $0$.  Angular momentum conservation dictates that only the 
helicity-$0$ state of the $[1^{--}]^{\mathbf 8}$ can be produced at 
rest or low momentum in $gg$ fusion.  When the helicity-$0$ 
$[1^{--}]^{\mathbf 8}$ state decays to the $[2^{++}]^{\mathbf 1}\ 
\chi_{2}$, it can only yield the \c2\ in the helicity $\pm 1$ state.
This completes our argument for point (2) in our list.

Our principal results, eq.~(\ref{unpol}) and (\ref{ang}), follow 
directly from the helicity selection rules (1) and (2) 
and the angular distributions of 
eq.~(\ref{dist}).  The attractive feature of this analysis is that the 
reaction mechanism can be determined by the study of the \c2\ decay 
angular distribution in the unpolarized case (eq.~(\ref{unpol})), 
after which the gluon distribution in the proton can be determined by 
analysis of the polarization asymmetry of the decay angular 
distribution (eq.~(\ref{ang})).
 

\end{document}